\documentclass[aps,prl,twocolumn,amsmath,amssymb]{revtex4}
\usepackage{graphicx}
\begin{document}

\title{Generation of Atomic Cluster States through the Cavity Input-Output Process}
\author{Jaeyoon Cho}
\affiliation{Division of Optical Metrology, Korea Research Institute of Standards and Science, Daejeon 305-340, Korea}
\author{Hai-Woong Lee}
\affiliation{Department of Physics, Korea Advanced Institute of Science and Technology, Daejeon 305-701, Korea}
\date{\today}
\begin{abstract}
We propose a scheme to implement a two-qubit controlled-phase gate for single atomic qubits, which works in principle with nearly ideal success probability and fidelity. Our scheme is based on the cavity input-output process and the single photon polarization measurement. We show that, even with the practical imperfections such as atomic spontaneous emission, weak atom-cavity coupling, violation of the Lamb-Dicke condition, cavity photon loss, and detection inefficiency, the proposed gate is feasible for generation of a cluster state in that it meets the scalability criterion and it operates in a conclusive manner. We demonstrate a simple and efficient process to generate a cluster state with our high probabilistic entangling gate.
\end{abstract}
\pacs{pacs}
\maketitle

\newcommand{\bra}[1]{\left<#1\right|}
\newcommand{\ket}[1]{\left|#1\right>}
\newcommand{\abs}[1]{\left|#1\right|}

The one-way quantum computation \cite{br01,rb01,rbb03,n04,nd04} has opened up a new paradigm for constructing reliable quantum computers. In their pioneering works \cite{br01,rb01}, Raussendorf and Briegel showed that preparation of a particular entangled state, called a cluster state, accompanied with local single-qubit measurements is sufficient for simulating any arbitrary quantum logic operations. Therefore, experimental or intrinsic difficulties in performing two-qubit operations can be substituted with (possibly probabilistic) generation of an entangled state. Especially, Nielsen showed that the resource overhead of a conventional linear optics quantum computer \cite{klm01} is drastically decreased by combining it with the idea of the one-way quantum computation \cite{n04}.

A cluster state can be visualized as a collection of qubits and lines connecting them. In order to generate a cluster state systematically, one first initializes each qubit in state $\ket{+}=\frac{1}{\sqrt2}(\ket{0}+\ket{1})$, where $\ket{0}$ and $\ket{1}$ are the computational basis states, and then performs controlled-phase operations between every neighboring qubits connected by the lines. In some previous works \cite{bk04,dr05,br05}, it was shown that in principle there is no threshold value of $p$ required for efficient generation of a cluster state, where $p$ is the success probability of each controlled-phase operation. For a reasonable computational overhead, however, a high success probability $p$ should be attained.

In the present work, we propose a scheme to implement a two-qubit controlled-phase gate for single atomic qubits, which works in principle with nearly ideal success probability and fidelity. The proposed entangling gate is suitable for the systematic generation of a cluster state described above for two reasons. The first is that it works between two individually trapped atoms, thus it meets the scalability criterion. Since a large number of qubits should be entangled in a cluster state to perform a nontrivial quantum computation, entangling gates which work only inside a single trapping structure \cite{cz95,pgcz95,pw02,ysy03} can not be used directly for our goal. The second is that, in contrast to other scalable two-qubit gate schemes \cite{cz20,kmw02,xlgy04}, it operates in a conclusive manner even in the practical situation. Even if the success probability decreases due to the experimental imperfection, one can still detect whether the operation has succeeded or not, and in case it has succeeded, the fidelity is very high \cite{bk04,lbk05}. We demonstrate how a cluster state of an arbitrary configuration can be generated with our high probabilistic entangling gate.


\begin{figure}
\centerline{\includegraphics[width=0.3\textwidth]{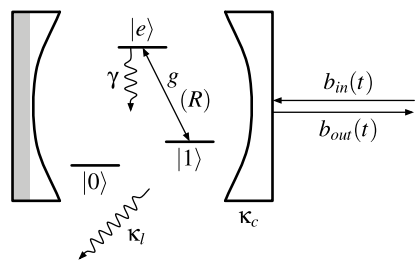}}
\caption{The setup for the basic building block. A qubit is encoded in two ground levels $\ket0$ and $\ket1$ of a 3-level atom trapped in an one-sided optical cavity. The transition between states $\ket1$ and $\ket{e}$ is coupled resonantly to the right-circularly polarized mode of the cavity with coupling rate $g$. $\gamma$ denotes atomic spontaneous emission rate. The cavity photon is either transmitted through the cavity mirror with rate $\kappa_c$ or lost with rate $\kappa_l$. $b_{in}(t)$ and $b_{out}(t)$ denote the input and the output field operators, respectively.}
\label{fig:setup}
\end{figure}

Fig.~\ref{fig:setup} shows the setup for the basic building block of our scheme. A qubit is encoded in two ground levels $\ket0$ and $\ket1$ of a 3-level atom, which is trapped in an one-sided optical cavity. The transition between states $\ket{1}$ and $\ket{e}$ is coupled resonantly to the right-circularly polarized mode of the cavity with coupling rate $g$, and state $\ket{0}$ is decoupled from the cavity field. We consider two kinds of transition channels for the cavity photon. The first one is the cavity decay due to transmission through the cavity mirror, whose rate is $\kappa_c$. Every other unwanted photon losses, such as cavity absorption and scattering, are characterized by the overall loss rate $\kappa_l$. For the gate operation, we will inject a photon into the cavity and observe the output photon along the cavity decay channel, and postselect those cases in which a photon is detected. The evolution of the system, then, can be described by the non-Hermitian conditional Hamiltonian in the framework of the quantum trajectory method \cite{c93}. In the rotating frame, the conditional Hamiltonian of the system, without the cavity decay, can be written as
\begin{equation}
H_s=-i\frac{\gamma}{2}\ket{e}\bra{e}+g(a\ket{e}\bra{1}+a^\dagger\ket{1}\bra{e})-i\frac{\kappa_l}{2} a^\dagger a,
\end{equation}
where $\gamma$ and $a$ denote the atomic spontaneous emission rate and the annihilation operator for the right-circularly polarized mode of the cavity, respectively. Taking into account the coupling through the cavity decay channel, the system is fully specified by the boundary condition
\begin{equation}
b_{out}(t)=b_{in}(t)+\sqrt{\kappa_c}a(t),
\label{eq:03141}
\end{equation}
and the quantum Langevin equation
\begin{equation}
\begin{split}
\dot{s}=-i(sH_s-H_s^\dagger s)&-[s,a^\dagger]\left(\frac{\kappa_c}{2}a+\sqrt{\kappa_c}b_{in}(t)\right)\\
&+[s,a]\left(\frac{\kappa_c}{2}a^\dagger+\sqrt{\kappa_c}b_{in}^\dagger(t)\right),
\end{split}
\label{eq:03142}
\end{equation}
where $s$ is an arbitrary system operator, and $b_{in}(t)$($b_{out}(t)$) is the input(output) field operator \cite{gc85}.

Suppose the atom is initially prepared in its ground state. When a photon is reflected from the cavity, its pulse shape would be changed due to the interaction with the atom-cavity system. In particular, when both the adiabatic condition ($\abs{\frac{\dot{s}}{s}}\ll\kappa_c,g$) and the strong atom-cavity coupling condition ($g\gg\kappa_c,\gamma$) are satisfied, the system only acquires a conditional phase shift with a good approximation \cite{dk04}. If the atom is in state $\ket{1}$ and a right-circularly polarized photon is incident, the system acquires no phase shift. Otherwise, i.e., if the photon does not see the atom, the system acquires a phase shift of $\pi$. Accordingly, in this regime, the simple setup of Fig.~\ref{fig:setup} serves as a controlled-phase gate between a photonic qubit and an atomic qubit.

Before introducing the complete scheme, let us investigate this building block in more detail taking into account various aspects of practical imperfections. We assume the atom is trapped in a harmonic potential. Since the cavity field varies spatially along the cavity axis, the harmonic motion of the atom leads to time variation of the atom-cavity coupling rate. With an assumption that the gate operates outside the Lamb-Dicke condition, we model the time dependence of the atom-cavity coupling rate as $g(t)=g_0\cos\left(\frac{\pi}{3}\sin\left(\frac{2\pi t}{T_g}+\phi\right)\right)$, where $T_g$ denotes the period of the atomic motion and $\phi$ is an arbitrary phase. Here, we have allowed the coupling rate to vary between $g_0/2$ and $g_0$ in accordance with a typical cavity QED experiment \cite{emky01}. The pulse shape of the input photon is assumed to be $f_{in}(t)=\left[T_f \cosh\left(\frac{2t}{T_f}\right)\right]^{-1}$, which is normalized as $\int\abs{f(t)}^2=1$. Here, $T_f$ denotes the pulse width. We define $P$ as the success probability that a photon is detected, which is identical to the probability that no photon is lost by the atomic spontaneous emission (with rate $\gamma$) or the unwanted cavity photon loss (with rate $\kappa_l$). Since we postselect those cases in which a photon is detected, the pulse shape $f_{out}(t)$ of the output photon can be regarded to be normalized as $\int\abs{f_{out}(t)}^2=1$, and the fidelity $F$ between the two pulses is given by $F=\abs{\int f_{in}^*(t)f_{out}(t)dt}$. All of the values above can be obtained on the basis of the cavity input-output formulae (\ref{eq:03141}) and (\ref{eq:03142}). 


\begin{figure}
\centerline{\includegraphics[width=0.45\textwidth]{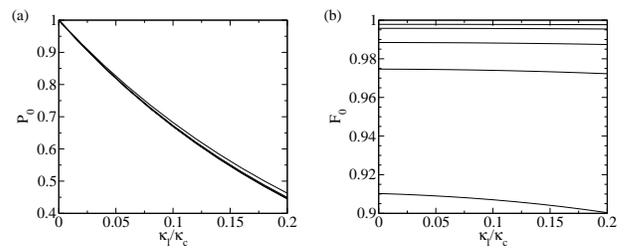}}
\caption{When the photon does not interact with the atom, (a) the success probability $P_0$ and (b) the fidelity $F_0$ with respect to $\kappa_l$ for $\kappa_c T_f=\{10,20,30,50,70\}$. In (b), the upper curve is obtained for the longer $T_f$ in order.}
\label{fig:graph1}
\end{figure}


\begin{figure}
\centerline{\includegraphics[width=0.45\textwidth]{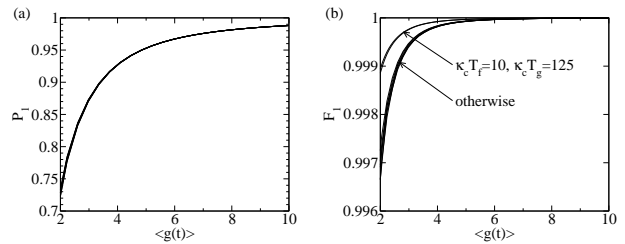}}
\caption{When the photon interacts with the atom, (a) the success probability $P_1$ and (b) the fidelity $F_1$ with respect to the average atom-cavity coupling rate $\left<g(t)\right>$ for every combinations of parameter sets: $\kappa_c T_f=\{10,50\}$, $\kappa_c T_g=\{50,125\}$, and $\kappa_l/\kappa_c=\{0,0.2\}$.}
\label{fig:graph2}
\end{figure}

Let us first consider a case in which a photon reflects from a bare cavity. Let $P_0$ and $F_0$ be the success probability and the fidelity in this case, respectively. In Fig.~\ref{fig:graph1}, we plot (a) $P_0$s and (b) $F_0$s with respect to $\kappa_l$ varying the pulse width: $\kappa_c T_f=\left\{10,20,30,50,70\right\}$. Fig.~\ref{fig:graph1}(a) shows the success probability is determined dominantly by $\kappa_l$: $P_0$ decreases as $\kappa_l$ increases. In Fig.~\ref{fig:graph1}(b), the upper curve is obtained for the longer $T_f$ in order. This behavior is originated from the fact that the adiabatic condition is satisfied more strongly with the longer pulse width. When $\kappa_c T_f\gtrsim50$, the attained fidelity is found to be very close to the ideal value ($F_0>0.995$) regardless of the cavity photon loss. Our numerical calculations indicate that, in every cases, the acquired phase shift is exactly $\pi$. Secondly, we consider another case in which a right-circularly polarized photon reflects from the cavity while the atom is prepared in state $\ket{1}$. In this case, due to the interaction between the photon and the atom, the reflection occurs in a different manner. Let $P_1$ and $F_1$ be the success probability and the fidelity in this case, respectively. In Fig.~\ref{fig:graph2}, we plot (a) $P_1$s and (b) $F_1$s, which have been averaged over $\phi$, with respect to the average atom-cavity coupling rate $\left<g(t)\right>$ for every combinations of parameter sets: $\kappa_c T_f=\left\{10,50\right\}$, $\kappa_c T_g=\left\{50,125\right\}$, and $k_l/\kappa_c=\left\{0,0.2\right\}$. Here, we have assumed $\gamma=\kappa_c$. In this case, our numerical simulations indicate that the cavity photon is hardly created. The photon loss is thus dominated by the atomic spontaneous emission. Accordingly, both $P_1$ and $F_1$ are determined dominantly by the atom-cavity coupling rate, which is why each curve in Fig.~\ref{fig:graph2} is hardly distinguishable. Fig.~\ref{fig:graph2}(b) shows that the fidelity is very close to the ideal value even in the weak atom-cavity coupling regime. The acquired phase is found to be exactly zero.

A remarkable point of the above numerical results is that, though the success probability could decrease due to the unavoidable photon loss, the fidelity remains very high in most parametric regimes we have considered. From now on, let us assume $F_0=F_1=1$ for simplicity. In order to demonstrate that the setup of Fig.~\ref{fig:setup} serves as a controlled-phase gate, suppose a photon in state $\frac{1}{\sqrt2}(\ket{L}+\ket{R})$, where $\ket{L}$($\ket{R}$) denotes a left-(right-)circularly polarized photon, is reflected from the cavity while the atom is in state $\frac{1}{\sqrt2}(\ket{0}+\ket{1})$. A straightforward calculation yields the success probability $P=\frac{P_0}{4}(3+r)$ and the fidelity $F=\frac{3+\sqrt{r}}{2\sqrt{3+r}}$, where we have defined $r\equiv P_1/P_0$. The resulting entangled state can be written as $\frac{1}{\sqrt{3+r}}(\ket0\ket{L}+\ket0\ket{R}+\ket1\ket{L}-\sqrt{r}\ket1\ket{R})$ up to a global phase.


\begin{figure}
\centerline{\includegraphics[width=0.3\textwidth]{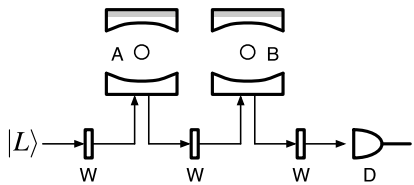}}
\caption{Controlled-phase gate between atom A and atom B. Each W represents a $\lambda/4$-plate and D represents a polarization detector. For the gate operation, a left-circularly polarized single photon is injected from left and the polarization of the output photon is measured at the detector.}
\label{fig:gate}
\end{figure}


\begin{figure}[b]
\centerline{\includegraphics[width=0.45\textwidth]{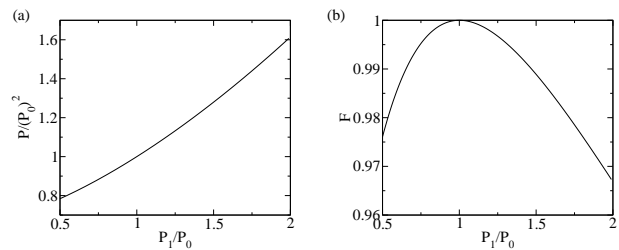}}
\caption{(a) The success probability $P$ and (b) the fidelity $F$ of the controlled-phase gate with respect to $r=P_1/P_0$.}
\label{fig:graph3}
\end{figure}

Now, the building block in Fig.~\ref{fig:setup} can be exploited for our goal. Fig.~\ref{fig:gate} shows the controlled-phase gate between two atoms A and B. Each W in the figure represents a $\lambda/4$-plate that converts the basis of a single-photon qubit between $\left\{\ket{L},\ket{R}\right\}$ and $\left\{\frac{1}{\sqrt2}(\ket{L}\pm\ket{R})\right\}$. Initially, each atom is prepared in state $\frac{1}{\sqrt2}(\ket0+\ket1)$. For the gate operation, a single photon in state $\ket{L}$ is injected from left and the polarization of the output photon is measured at the detector. From a straightforward algebra, one can find that a photon in state $\ket{L}$ is detected with probability $P_L=\frac{P_0^2}{32}\left[r^2+2r+4(r-1)\sqrt{r}+13\right]$, while a photon in state $\ket{R}$ with probability $P_R=\frac{P_0^2}{32}(r+3)^2$. In the former case, the final state becomes
\begin{equation}
\begin{split}
\ket{\Psi_L}=\frac{P_0}{\sqrt{8P_L}}&\left[\ket0_A\ket0_B+\ket0_A\ket1_B+\ket1_A\ket0_B\right.\\
&\left.-\frac{r+2\sqrt{r}-1}{2}\ket1_A\ket1_B\right],
\end{split}
\end{equation}
and in the latter case,
\begin{equation}
\begin{split}
\ket{\Psi_R}=\frac{P_0}{\sqrt{8P_R}}&\left[\ket0_A\ket0_B+\ket0_A\ket1_B-\sqrt{r}\ket1_A\ket0_B\right.\\
&\left.+\frac{r+1}{2}\ket1_A\ket1_B\right],
\end{split}
\end{equation}
which can be converted to the desired entangled state by applying a Pauli operator $\sigma_x$ on atom B. In Fig.~\ref{fig:graph3}, we plot (a) the success probability $P=P_L+P_R$ and (b) the average fidelity $F$ with respect to $r\equiv P_1/P_0$. Since a photon passes through two cavities in order, the success probability is basically second order in $P_0$ and $P_1$. The fidelity is found to be very high regardless of the success probability. In particular, when $P_0\simeq P_1$, the attained fidelity is as high as 1. An interesting property of the gate is that the fidelity would be decreased as the atom-cavity coupling rate is increased. In order to get an optimal fidelity, one first increase $F_0$ by increasing the pulse width as shown in Fig.~\ref{fig:graph1}(b), and then adjust $\kappa_l$ and $\left<g(t)\right>$ to have $P_0=P_1$. For a typical cavity decay rate $\kappa_c/2\pi=4~\mathrm{MHz}$ \cite{emky01}, one gets $F_0>0.995$ with $T_f=50/\kappa_c\simeq2~\mathrm{\mu s}$. We note that the success probability could decrease further due to photon losses at other parts of the setup in Fig.~\ref{fig:gate}, such as optical components, optical paths, and the detector. Even in those cases, the fidelity is not affected as long as the losses are polarization independent and dark counts are neglected. 


\begin{figure}
\centerline{\includegraphics[width=0.35\textwidth]{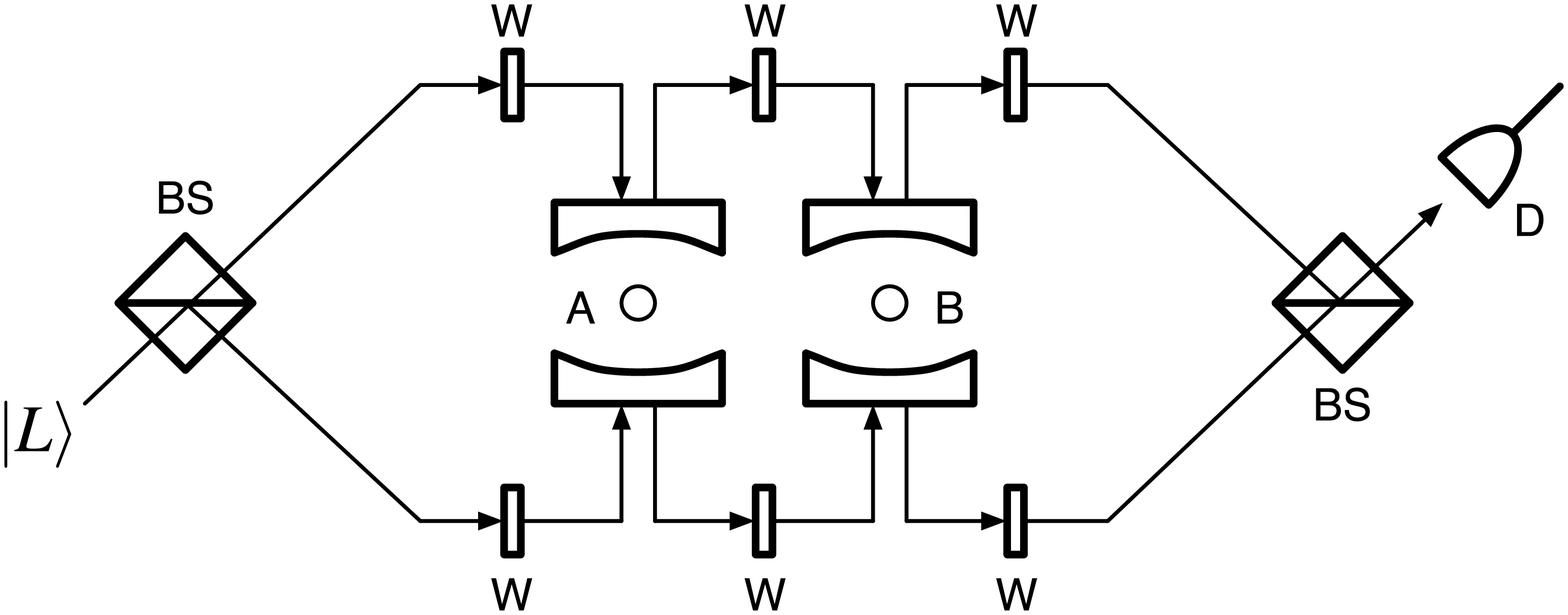}}
\caption{A modified version of the controlled-phase gate to take advantage of two-sided cavities. Each BS represents a 50:50 beam splitter.}
\label{fig:final}
\end{figure}

There is still room for improvement by which one can replace each one-sided cavity in Fig.~\ref{fig:gate} with a two-sided cavity. Let us assume both cavity mirrors have the same decay rate $\kappa_c'$. One can easily show that the cavity input-output formulae (\ref{eq:03141}) and (\ref{eq:03142}) as well as the commutation relations $\left[b_{in}^\dagger(t),b_{in}(t')\right]=\left[b_{out}^\dagger(t),b_{out}(t')\right]=\delta(t-t')$ are preserved by substituting as
\begin{equation}
\begin{split}
b_{in}(t)&\rightarrow\frac{1}{\sqrt2}\left[b_{in}^{(1)}(t)+b_{in}^{(2)}(t)\right],\\
b_{out}(t)&\rightarrow\frac{1}{\sqrt2}\left[b_{out}^{(1)}(t)+b_{out}^{(2)}(t)\right],\\
\kappa_c&\rightarrow2\kappa_c',
\end{split}
\end{equation}
where two cavity decay channels are represented by superscripts $(1)$ and $(2)$, respectively. The setup of Fig.~\ref{fig:final} thus works in the same fashion as that of Fig.~\ref{fig:gate} with an effective cavity decay rate $2\kappa_c'$, where each beam splitter is of 50:50 type.

Finally, we demonstrate how our controlled-phase gate is directly used to generate a cluster state. Here, we assume the gate works with success probability $P>2/3$. In this case, one can take a simple add-on strategy to generate a cluster state of an arbitrary configuration. In order to show this, let us denote by $\ket{\Psi_n}$ the 1D cluster state of $n$ qubits, and express $\ket{\Psi_{n-2}}$ as
\begin{equation}
\ket{\Psi_{n-2}}=\ket{\phi_0}_{n-3}\ket{0}_{n-2}+\ket{\phi_1}_{n-3}\ket{1}_{n-2},
\end{equation}
where $\ket{i}_{n-2}$ denotes the state of the $(n-2)$th qubit and $\ket{\phi_i}_{n-3}$ denotes the relevant terms for the other $(n-3)$ qubits. It is easily verified that $\ket{\Psi_n}$ can be written as
\begin{equation}
\begin{split}
\ket{\Psi_n}&=\frac{1}{\sqrt2}\ket{\phi_0}_{n-3}\ket{0}_{n-2}\left(\ket{0}_{n-1}\ket{+}_n+\ket{1}_{n-1}\ket{-}_n\right)\\
&+\frac{1}{\sqrt2}\ket{\phi_1}_{n-3}\ket{1}_{n-2}\left(\ket{0}_{n-1}\ket{+}_n-\ket{1}_{n-1}\ket{-}_n\right),
\end{split}
\label{eq:04041}
\end{equation}
where $\ket{\pm}=\frac{1}{\sqrt2}(\ket0\pm\ket1)$. In order to generate $\ket{\Psi_{n+1}}$, one simply attach a qubit in state $\ket+$ to $\ket{\Psi_n}$ by performing a controlled-phase operation. If the operation succeeds, one gets $\ket{\Psi_{n+1}}$. If it fails, however, since $n$th qubit is measured in an arbitrary basis, the state~(\ref{eq:04041}) becomes a mixed state
\begin{equation}
\begin{split}
\rho_{n-1}^f&=\frac12\left(\ket{\phi_0}_{n-3}\ket{0}_{n-2}+\ket{\phi_1}_{n-3}\ket{1}_{n-2}\right)\ket{0}_{n-1}\bra{\cdots}\\
&+\frac12\left(\ket{\phi_0}_{n-3}\ket{0}_{n-2}-\ket{\phi_1}_{n-3}\ket{1}_{n-2}\right)\ket{1}_{n-1}\bra{\cdots}.
\end{split}
\end{equation}
From this expression, it is apparent that $\ket{\Psi_{n-2}}$ can be recovered from $\rho_{n-1}^f$ by measuring the $(n-1)$th qubit in  the computational basis and performing an appropriate unitary operation on the $(n-2)$th qubit according to the measurement result. In other words, when an add-on process fails, only two qubits are lost. The average number of qubits attached by $m$ entangling operations is thus $(3P-2)m$, which grows on average if $P>2/3$. In the same fashion, it is also shown that if the $i$th qubit of $\ket{\Psi_n}$ $(i<n)$ is measured in an arbitrary basis, one can recover two 1D cluster states $\ket{\Psi_{i-2}}$ and $\ket{\Psi_{n-i-1}}$ up to appropriate local unitary operations by measuring both the $(i-1)$th and the $(i+1)$th qubits. We can thus connect two 1D cluster states by performing controlled-phase operations to form a cross-shaped 2D cluster state. Though a failure of the entangling operation would break them into four 1D cluster states, they can be connected into two 1D cluster states as shown above, and then be used to form a 2D cluster state again. By repeating these procedures, one can generate a cluster state of an arbitrary configuration.

In summary, we have proposed a contolled-phase gate which operates between two distant atoms each trapped in an optical cavity, and have shown that the proposed gate is feasible for generation of a cluster state. In particular, the gate has no theoretical bound on the attainable success probability while it achieves a very high fidelity even with the considerable imperfections.

This research was supported by a Grant from the Ministry of Science and Technology (MOST) of Korea and from Korea Research Institute of Standards and Science (KRISS).


\end{document}